\documentclass{hep99}
\usepackage{epsfig}

\begin{document}

\title{Object oriented data analysis in ALEPH}

\author{Giuseppe Bagliesi}

\address{Istituto Nazionale di Fisica Nucleare, S. Piero a Grado, Pisa, Italy\\
E-mail: {\tt Giuseppe.Bagliesi@cern.ch}}

\abstract{This article describes the status of the ALPHA$^{++}$ project of the ALEPH collaboration. The 
ALEPH data have been converted from Fortran data structures (BOS banks) into C$^{++}$ objects
and stored in a object oriented database (Objectivity/DB), using tools provided by the RD45 
collaboration and the LHC$^{++}$ software project at CERN. A description of the database setup
and of a preliminary version of an object oriented analysis program is given.
} 

\maketitle


\section{Introduction}
Object oriented programming is becoming the new paradigm for software development in
high energy physics, in particular in view of the upcoming, very demanding LHC project
and its experiments. In order to provide an analysis platform and OO application software,
the LHC$^{++}$ project has been set up at CERN\cite{lhc++}.
Furthermore, commercial products for the storage of persistent objects in so called object
databases are being evaluated by the RD45 collaboration\cite{rd45}.
The current candidate for a large scale application at CERN is Objectivity/DB\cite{objy}. 
In ALEPH a project called ALPHA$^{++}$\cite{alpha++} has been set up, with the following goals:

\begin{itemize}
\item convert the ALEPH data from a Fortran (BOS\cite{bos}) bank style into persistent objects
and write them to an object database (Objectivity/DB)
\item rewrite a mini-version of the ALEPH analysis package ALPHA\cite{alpha} in an object oriented
computing language (C$^{++}$), based on the object database
\item compare the standard and OO performance with regard to the efficient access of the data 
and their manipulation
\item test the software engineered by the RD45 and LHC$^{++}$ projects
\item provide input and experience for a possible archiving of ALEPH data
\item give an opportunity to learn OO analysis, design and programming.  
\end{itemize} 

In this report we present the structure of the ALEPH object database and a preliminary version
of the analysis program.

\section{The ALEPH data structure and its conversion to objects}

\begin{figure}
\begin{center}
\epsfig{file=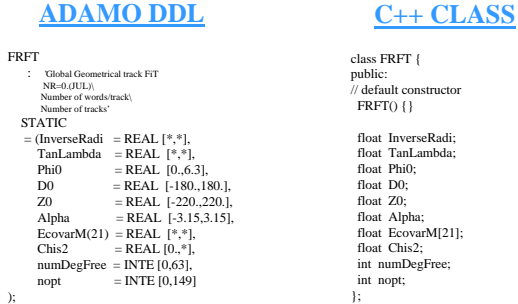,height=4cm}
\end{center}
\caption{Correspondence between the ADAMO description of the track bank FRFT and its C$^{++}$ implementation }
\label{fig:fig1}
\end{figure}

ALEPH uses the Fortran based BOS system for the memory management. The event data are kept in memory 
in a large array which is globally accessible as a common block. The data are organised in so called
{\em banks}, and the data definition language (DDL) for these banks is provided by ADAMO\cite{adamo}.
The ADAMO package offers a conversion to C header files, which facilitates the automatic conversion
of the ALEPH DDL to C$^{++}$ classes, or to be more precise, to the ddl-files needed for the setup
of an Objectivity/DB database. It has been decided to perform a one-to-one conversion of all the
 relevant banks stored on an ALEPH DST (Data Summary Tape), which can be read by the analysis 
program ALPHA. A simple C$^{++}$ tool has been developed which allows for the automatic conversion
of the ADAMO DDL of all 173 relevant banks into C$^{++}$ classes. An example of the correspondence
between the ADAMO description of the track bank FRFT and its C$^{++}$ implementation is given in 
Fig.\ref{fig:fig1}.

\begin{figure}
\begin{center}
\epsfig{file=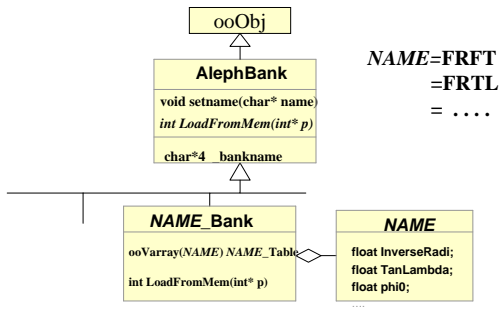,height=4cm}
\end{center}
\caption{The DDL structure of the ALEPH object database} 
\label{fig:fig2}
\end{figure}

The general DDL structure as implemented in the Objectivity database is outlined in Fig.\ref{fig:fig2}.

A class {\em AlephBank} serves as a base class for all the ALEPH banks. Persistency is obtained
by inheritance from the Objectivity class ooObj. The class {\em AlephBank} stores the name of each
bank, and defines the interface for the reading from and writing to memory of the bank contents.
Each ALEPH bank is implemented as a class {\em NAME\_Bank}, where {\em NAME} is the BOS name of the
bank, such as FRFT. This class {\em NAME\_Bank} contains the instances of the class {\em NAME}.

\section{The ALEPH object database}
\begin{figure}
\begin{center}
\epsfig{file=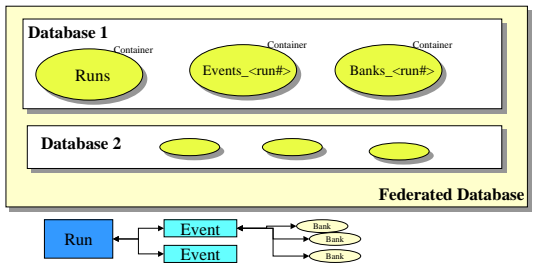,width=7.0cm}
\end{center}
\caption{The database structure of the ALEPH object database} 
\label{fig:fig3}
\end{figure}

The database structure of the ALEPH object database is reproduced 
in Fig.\ref{fig:fig3}. At the basis there is a federated database called ALEPHDB,
which contains several databases. These databases either contain real data from
different data taking periods with or without pre-classification of events,
or Monte Carlo events from fully simulated hadronic Z decays.

Every database has a container holding {\em Run} objects. These objects 
store the run number and have bi-directional links to all the event objects 
of the corresponding run and to banks which store global run condition data. 
Furthermore, for every run there is a container with the event objects
of this run and an additional container with all the bank objects per run.
The event objects store the event number and the classification bit of the
 event, and have bi-directional links to all the banks of the event.
At the moment about eight Gbyte of real and Monte Carlo data are stored in the
federated database. The size of an hadronic Z decay event is $\approx 114KB$ in the old 
EPIO format and $\approx 145KB$ in the object database. The overhead arises mainly
from the relationships which are stored in the object database.

\section{Analysis Program}
 \label{program}
The ALPHA$^{++}$ analysis program is based on the structure of ALPHA. Two basic objects are defined:
 the $Tracks$ and the $Vertices$.

A $Track$ can be a charged track reconstructed in the TPC, a photon identified in the ECAL 
or a general Energy Flow Object.
The common attributed of a $Track$ are: the total momentum with its components, the energy, 
the mass and the charge. A $Track$ can also have more specific attributes depending if it is a charged track, a photon or a Energy Flow Object. 

\begin{figure}
  \begin{center}
    \epsfig{file=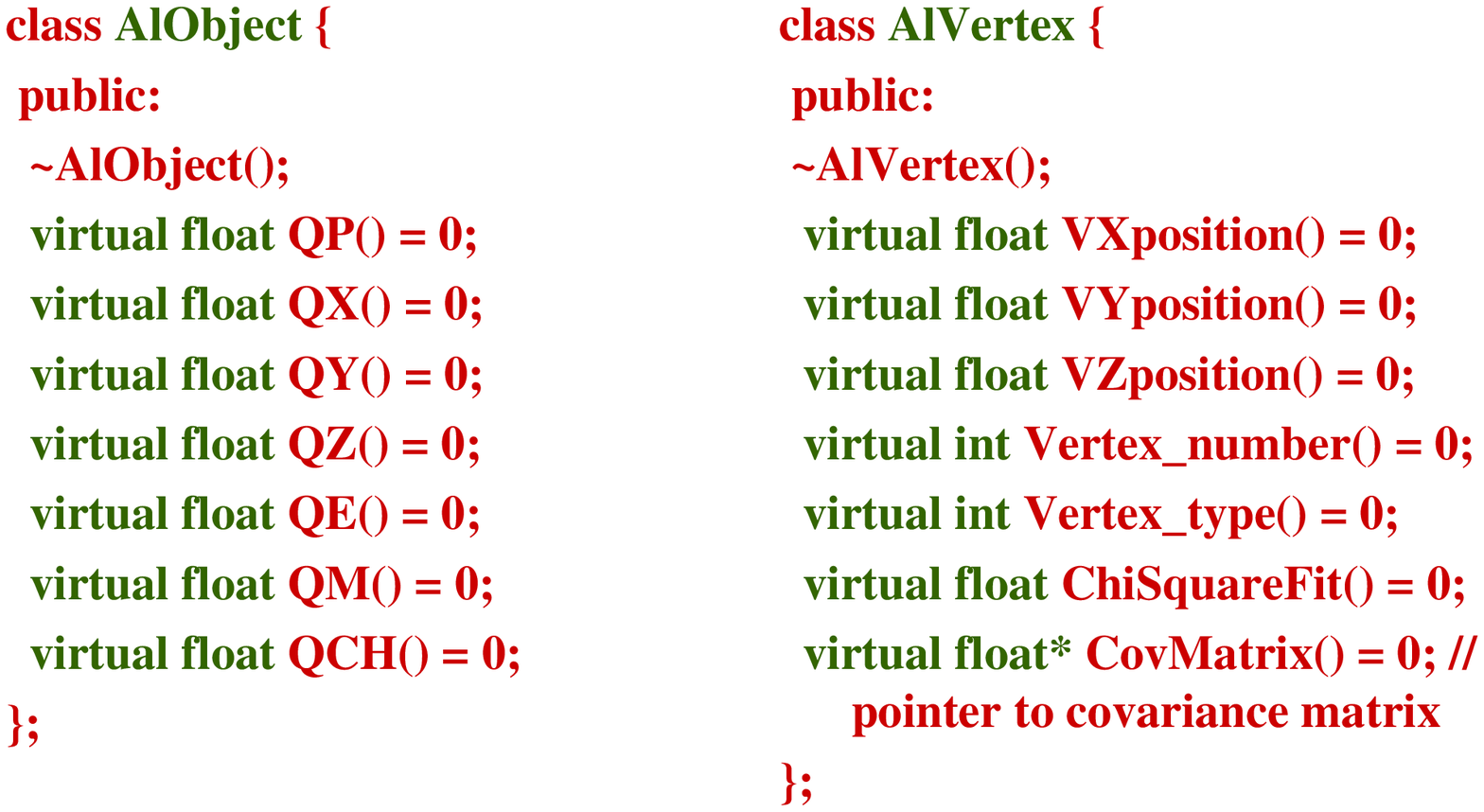,width=7.0cm}
    \caption{The $AlObject$ and $AlVertex$ abstract classes interface \label{alpha:abstract}}
  \end{center}
\end{figure}

\begin{figure}
  \begin{center}
    \epsfig{file=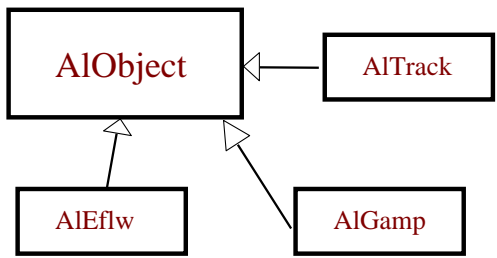,width=6.5cm}
    \caption{The $AlObject$ inheritance structure \label{alpha:Objects}}
  \end{center}
\end{figure}

The previously described structure has been implemented in ALPHA$^{++}$ with an abstract class $AlObject$
 (Fig.\ref{alpha:abstract}) with the common 
attributes of a $Track$ defined as purely virtual member functions. 
The concrete transient classes $Altrack$ (charged tracks),
 $AlEflw$ (Energy Flow) and $AlGamp$ (photons) are defined by inheritance from $AlObject$ 
(Fig.\ref{alpha:Objects}).

\begin{figure}
  \begin{center}
    \epsfig{file=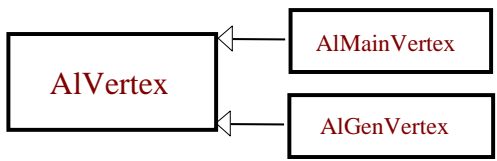,width=6.5cm}
    \caption{The $AlVertex$ inheritance structure \label{alpha:vertex}}
  \end{center}
\end{figure}

The same principles apply also in the definition of the $Vertices$ (Fig.\ref{alpha:abstract}). 
There is an abstract class $AlVertex$
 with the basic attributes, and two concrete transient classes $AlMainVertex$ 
holding the position of the interaction point,
 and $AlGenVertex$ for the secondary vertices found by the reconstruction program 
(Fig.\ref{alpha:vertex}).

The implementation of the previously mentioned
 transient classes in ALPHA$^{++}$ makes use of the ALPHA internal banks 
QVEC (containing the $Tracks$) and QVRT (containing the $Vertices$). 
This is done in the following steps:
\begin{itemize}
\item For each event the relevant persistent classes from the Objectivity database are read and the corresponding BOS banks in the fortran BOS common are filled;
\item the fortran subroutines which starting from the BOS banks construct the internal QVEC and QVRT data structures are called from C$^{++}$;
\item the concrete C$^{++}$ classes (like $AlTrack$, $AlMainVertex$ ...) are instantiated using the data contained in QVEC and QVRT;
\item for each event the transient class $AlphaBanks$ containing all the instances of 
the previous classes and giving access to them through member functions is instantiated.
 \end{itemize}
 
The choice to call FORTRAN from C$^{++}$ in the analysis program is due to the fact that there are many algorithms
in ALPHA which, on a short time scale, are not possible to develop in C$^{++}$. 
Therefore our strategy is to have a FORTRAN ``wrapped''  analysis program already working and to use 
it as a basis to develop new C$^{++}$ code and algorithms.

\section{Preliminary performances studies}
In order to check the performances of the Objectivity/DB 
a very simple analysis program has been set up both in ALPHA and ALPHA$^{++}$. 
This program applies some cuts to hadronic Z decays events to 
preselect good events for the QCD analyses. In addition some histograms are filled.
The preliminary results are shown in Table~\ref{perf_table}.
We found that the analysis CPU time is negligible as well as the histogram filling time. The factor $\approx 2$
difference in CPU time per event between ALPHA$^{++}$ and ALPHA is due to the input/output operations performed
by the Objectivity/DB.

\begin{table}
\begin{center}
\begin{tabular}{lp{1.5cm}p{1.5cm}p{1.5cm}} 
\br
  &$t_{hadr}$ (sec/ev)&$t_{all}$ (sec/ev)& $t_{init}$ (sec) \\ 
\mr
 ALPHA   &  $15.1\cdot 10^{-3}$      & $1.9\cdot 10^{-3} $    & $1.48$ \\ 
 ALPHA$^{++}$ &  $29\cdot 10^{-3}$        & $2.6\cdot 10^{-3}$     & $1.75$ \\ 
\br
\end{tabular}
\end{center}
\caption{Preliminary results. The CPU time for Hadronic Events ($t_{hadr}$), 
for all triggered events ($t_{all}$)
and the job initialisation time ($t_{init}$) are 
given. The CPU time is in Alpha 8400 units, corresponding to 185 CERN units.}
\label{perf_table}
\end{table}
 
\section{Conclusions}
A project called ALPHA$^{++}$ has been set up with the ALEPH experiment in order to test
new object oriented approaches to data storage and analysis. The DDL structure of the object database
and a preliminary version of an OO analysis program have been presented. Future work will concentrate
on the further development of the analysis program as well as on detailed performances studies.

\end{document}